\documentstyle[12pt]{article}
\input psfig.sty
\def\simgt{\lower.5ex\hbox{$\; \buildrel > \over \sim \;$}}
\def\simlt{\lower.5ex\hbox{$\; \buildrel < \over \sim \;$}}

\def\Msun{$M_\odot$}
\def\Teff{$T_{\rm eff}$}

\def\dv{$\Delta V$}

\title{\bf The HR Diagram of Globular Clusters: Theorist' view(s)}
\author{F.~D'Antona \\
\vspace{1cm}\\
\normalsize{Osservatorio Astronomico di Roma, Italy}
\\}
\date{\mbox{}}
\begin{document}
\maketitle
\pagestyle{empty}
%
%
\def\bull{\vrule height .9ex width .8ex depth -.1ex}
\makeatletter
\def\ps@plain{\let\@mkboth\gobbletwo
\def\@oddhead{}\def\@oddfoot{\hfil\tiny\bull\quad
``The Galactic Halo~: from Globular Clusters to Field Stars'';
35$^{\mbox{\rm th}}$ Li\`ege\ Int.\ Astroph.\ Coll., 1999\quad\bull}%
\def\@evenhead{}\let\@evenfoot\@oddfoot}
\makeatother
%
%
\def\beginrefer{\section*{References}%
\begin{quotation}\mbox{}\par}
\def\refer#1\par{{\setlength{\parindent}{-\leftmargin}\indent#1\par}}
\def\endrefer{\end{quotation}}
%
%
{\noindent\small{\bf Abstract:} 
I list the characteristic features of Globular Cluster (GC) HR diagrams
which provide a complete test of the stellar evolution of low mass stars:
morphologies describing the different evolutionary phases, number ratios and
luminosity functions, which add quantitative information. Then I explore
the stage of todays' understanding of the classical distance scale
indicators (providing warnings against model construction which, underneath,
already is based on a choice of distance), and compare them to the
``new" indicators, such as the white dwarfs and the first and second kink of
the low main sequence. The classical and new distance indicators are still
subject to uncertainties due to tiny details of the theory, which are all of
the same order of magnitude, $\sim 0.25$mag and the absolute ages of GCs can
not be constrained to better than 10 -- 16 Gyr. However, most of recent
theoretical and observational results (including both classical and new
indicators) point more towards the lower range of ages (10-12) than to the
upper range (15-16).
 }
%
%
\section{Introduction}
It is now close to forty years that Globular Clusters (GCs) HR diagrams are
used to derive the age of the oldest stars of the Galaxy, but today they
provide a very complete test of the predictions of stellar evolution of low
mass stars. This is easily recognized by looking at the composite HR diagram
of the GC NGC 6397 shown in figure \ref{f1}, and illustrated in figure
\ref{f2} by recent stellar models. The diagram shows the core H--burning
phases (main sequence --MS-- and turnoff --TO-- regions, the H--shell burning
of the evolving mass ($M \simeq 0.8$\Msun) including the red giant branch
--RGB--, the Helium core burning phase (horizontal branch --HB--) and the
double shell burning phase (asymptotic giant branch --AGB--). The
Planetary Nebula phase is too short (only two planetaries are
 known in the
galactic GC system) but the white dwarf --WD-- cooling is well represented.
The MS ends at the top with the evolving stars, and at the bottom
with the lowest masses which are able to ignite hydrogen. The
characteristic {\it shape} of the MS well below the turnoff displays
characteristic details of the physics of the atmospheres and interiors of low
and very low mass stars (see, e.g. D'Antona 1995): it changes
slope twice, at a first, more luminous ``kink" (FK) at $M \sim 0.5$\Msun, and
at a second kink (SK), very close to the end of hydrogen burning structures.
So the HR diagrams provide {\it morphological} information and reference {\it
luminosity indicators} (TO, HB, WDs) which give constraints on the cluster
stars evolution. In addition the {\it number ratios} (luminosity functions
(LFs), ``clumps", gaps, ratio of HB to RGB number stars) add valuable
quantitative information to the morphology.

In particular, the LF of the turnoff and giant branch depends mostly on the
age of the system, while the mass function affects its unevolved part. The LF
below the TO presents a characteristic broad maximum, due to the functional
form of the mass -- luminosity relation, whose peak becomes dimmer when the
metal content increase (for a full description see, e.g., Silvestri et al.
1998).

In my opinion, the two fundamental questions which we would like to address
in a meeting on the comparison between GCs and halo field stars are the
following:
\begin{enumerate}
  \item is our knowledge sufficient to constrain the GC ages at a level at
which they can be interesting as indicators of the age of the Universe?
  \item are GC stars identical to their halo counterparts of similar
metallicity?
\end{enumerate}

The answers to these questions are simple and a bit unconfortable:
\begin{itemize}
  \item {\it Details of the theory} determine the precise absolute ages of
GCs: till now it is difficult to constrain the ages to better than from 10
to 16Gyr (but see later);
  \item The distance scale of GCs is necessary to know their ages. Its
calibration necessarily {\it relies on the hypothesis} that GC and halo stars
are strict relatives, unless we put all our faith on theoretical models only.
\end{itemize}

In the following I describe the ways in which we can try to determine the
distance scale of GCs and the hidden dangers in the playing of the
``isochrone fitting computer game", mainly those related to the use of the
giant branch location.
I will finally show how the new distance indicators emerging in the latest
years (WDs, FK and SK of the low MS) are consistent with the traditional
distance scales based on the fit of ground based photometry and on the HB
models.

\section{The ages paradigm}

The distance scale of GCs is the main key to their age determination. There
are many ``traditional" methods to derive this scale: they can be reduced to
the following list:
\begin{itemize}
  \item Approach based on distance indicators:
     \begin{itemize}
     \item Fitting of MS to local sample of subdwarfs;
     \item Fitting of the GCs RR Lyrae to RR Lyrae in the
Magellanic Clouds, distance of LMC calibrated through the Cepheids;
     \item Fitting of HB (or RR Lyrae) to local halo HB or RR Lyrae.
     \end{itemize}
  \item Purely theoretical approach:
     \begin{itemize}
     \item Fitting of HB (or RR Lyrae) luminosity to HB theoretical
models;
     \item Fitting of observed MS to theoretical MS (this implies a
match of the models colors);
     \item Fitting of the morphology of the HR diagram ($\delta (B-V)$
type methods)
     \end{itemize}
\end{itemize}

The recent, mainly HST based, observations which allow to reach dimmer and
dimmer luminosities have added new ways to determine the
distance, or at least to check it, which will be examined in sections
\ref{lowms}, \ref{secondk}\ and \ref{wd}, namely:
\begin{itemize}
   \item fitting of the location of the low MS, with attention to the
location of the FK;
   \item fitting of the MS region following the SK with the
local M subdwarfs sample;
   \item fitting of the WD sequence to models or disk counterparts.
\end{itemize}

I will not discuss the approach based on the classical distance indicators,
which has known a renewed interest in these years, thanks to the impact of
the results from the Hipparcos satellite. The fitting to the local subdwarfs
is discussed in Reid 1997, Gratton et al. 1997, Pont et al. 1998 and
Chaboyer et al. 1998. The RR Lyrae calibration after Hipparcos data was
first
rediscussed by Feast and Catchpole (1997). Notice that, while most Hipparcos
results imply a more or less stringent confirmation of the so called ``long"
distance scale for GCs, the local RR Lyrae and HB stars give a much shorter
scale (Fernley et al. 1998) consistent with the previous statistical
parallax determination by Layden et al. 1996 --but see the recent approach by
Groenewegen and Salaris 1999.

I will mostly concentrate on the theoretical approach. Actually, none of the
theoretical methods has been ever used independently from the others, but a
sort of ``consistency" between the different aspects of the problems has
generally been looked for, including the observational distance indicators.
For instance, fitting the morphology (that is, the relative position of RGB
and TO) as an absolute method for the age determination has never been taken
seriously  (see later) but the consistency of the whole HR diagram locations
(as shown in figure 1b) has been more or less unconsciously taken as
selfevidence of an evolutionary scheme --and thus of a given range of
ages-- and especially in some comparisons with observations we have seen
mention of ``spectacular fit", or ``location and shape matched superbly by
isochrones" while the truth hidden below is very different.

In the following I will base part of my discussion on a personal
interpretation of the events which in recent years led to a revision of the
average age of GCs. A look at the literature in fact shows that the ages of
GCs quoted before or up to 1996-1997 (pre-Hipparcos) are in the range
13-18Gyr, while the most quoted ages of the years 1998-1999 are 10-14Gyr
(post-Hipparcos)\footnote{Two notable exceptions are the work by Salaris et
al. (1997) and the three papers by our group (Mazzitelli et al. 1995,
D'Antona et al. 1997, Caloi et al. 1997) which published post--Hipparcos
ages in the pre--Hipparcos years}.
Vandenberg et al. 1996 in fact give an age of $15.8 \pm 2$Gyr to M92, noting
that ``ages below 12 or above 20Gyr appear highly unlikely", and Chaboyer et
al. (1996) give an average age of $14.6 \pm 1.7$Gyr to the galactic GCs,
putting a 95\% lower bound at 12.1Gyr. The turning point seems to have been
the results from Hipparcos satellite, which on the one hand made the metal
poor subdwarf sequence more luminous (by no more than 0.1mag, the effect
being even lower according to some researchers) and on the other hand
contributed to raise the zero point of Cepheids' luminosity, leading thus to
confirm a larger luminosity of the RR Lyrae in the LMC. However, the
Hipparcos results alone {\it do not} justify the global shift of the average
age of GCs, which amounts to $\sim 4$Gyr (from 16 to 12Gyr). In my opinion,
Hipparcos has simply given more weight to the  the ``long" distance scale of
GCs, which already had some enphasis in the observational literature (Sandage
1993, Walker 1992).

What has been happening is schematized in a very naif way in figure
\ref{deltav}: the HB (or RR Lyrae) luminosity has been increased in recent
models due to the sum of two small effects (a slight increase in the core
mass at the helium flash -by about 0.01\Msun- and a slight increase due to
the improvement in the equation of state (EoS)). 
This has been the most important update in the
models, and has shifted the ages to at most a couple of Gyr smaller {\it with
respect to previous HB theoretical luminosity}\footnote{Some confusion was
however present in the literature up to 1997 as to the absolute visual
magnitudes corresponding to the $\log L/L_\odot$\ of the models (see e.g. the
display in the lower panel of figure 1 from Caloi et al. 1997), so that the
modification induced in HB models might appear more drastic in some authors'
comparisons.}.

At the same time, the TO luminosity corresponding to a given age has been
slighlty decreasing. The sum of these subtle effects is a good $\sim
0.27$mag of difference in the absolute TO location at a given age based on
old or new models, for a given \dv\ from the HB to the TO, and thus the net
effect is a reduction of 4Gyr in the age. The interpretation which figure 3
gives to the age decrease is not {\it unique}: other motivations for a more
or less substantial decrease in the age are found in the recent literature.
Pont et al. (1998), who do not revise the HB luminosity, remark however
a difference in the scale of the $V$\ bolometric corrections ($BC_V$)
between the models atmospheres employed by Bergbush and Vandenberg 1992, and
the most recent scale both by Kurucz and by Bell, amounting to $\sim$0.1mag,
and thus leading to a decrease in the age by $\sim 1.5$Gyr.

It is evident that the HB luminosity of the present models, which in the
end represent the most direct classical distance indicator, is in itself
still uncertain, unless we believe that we can trust our models at the level
of 0.01\Msun\ for the determination of the helium core mass at flash, and
that we know perfectly all the other pieces of input physics. At least,
discussion is still open on the helium core flash masses. Notice also that,
e.g., the most recent HB models (Caloi et al. 1997 versus Cassisi et al.
1998) do not agree on the $L_{HB}$\ at intermediate metallicity (for [M/H]
from $\sim$ --1.5 to $\sim -1$), and it is unclear why. The problem of GC
ages is linked to minute details of the input physics, and we can not exclude
an uncertainty of $\sim 0.25$mag in the theoretical determination of the HB
luminosity. So the real uncertainty on the age determination from the HB is
still of $\pm$several Gyr.

However, why the paradigm of the 15-16Gyr age was so difficult to be
abandoned? In my opinion a part of the answer is the following:
{\it  in the course of many years, the whole theoretical construction of the
GC HR diagram had been adjusted to be consistent with about that age, so that
is resulted very difficult to make drastic changes to that view}. This
interpretation becomes more clear by examining the relative location of TO
and RGB in stellar models.

The TO color location is discussed in section \ref{toums}. The input
physics may affect it at a level of $\sim 0.05$mag.
On the other hand, the RGB colors {\it heavily}
depend on the treatment of convection. By changing the ratio mixing length
to pressure scale heigth ($\alpha=l/H_p$) in the MLT formulation,
the color location of the RGB may vary by tenths of magnitude (e.g.
Vandenberg 1983). Although everybody knew that the $\alpha$\ choice was `ad
hoc', the ``old" distance scale had this interesting outcome: by chance it
had the additional bonus of giving a very good fit of the MS and RGB
locations, if the models employed the same $\alpha$\ parameter which fitted
the solar radius at the solar age (solar calibration of the mixing length).
In addition, the solar calibration was also in good agreement with the
location of the best known subdwarf Groombridge 1830.
It was perhaps necessary to add some very small adjustment
of colors, but the reproduction of the GC morphologies was indeed very good.
The best example of this procedure is given by Bergbush and Vandenberg 1992:
they show how they calibrate their color-\Teff\ relations (based on quite
good model atmospheres) to provide a ``consistent" picture for all
metallicities. They also {\it explicitly state} that the transformations they
adopt are OK for their own models, and that different adjustments might be
required by other models. In spite of being very careful, the procedure
adopted in Bergbush and Vandenberg 1992 (but also by others) 
{\it implicitly hides both
the choice of the distance scale (and thus the resulting 15Gyr or so) and the
choice of the convection model}, as I will now clarify.
The fortuitous coincidence between the RG location in population II
models with a solar calibrated $\alpha$ led researchers to postpone the
problem of a better understanding of superadiabatic convection\footnote{This
``canonical" assumption was abandoned only in the Mazzitelli et al. 1995
paper, treating convection according to the Canuto and Mazzitelli (1991)
model, and in Chieffi et al. (1995), who propose a calibration of $\alpha$\
as a function of the cluster metallicity. This latter paper puts clearly into
evidence that no predictions can be made on the RGB location on the basis
of MLT models.}. After 1997 the -even small- change of distance scale implied
by the new HB models and by the Hipparcos subdwarfs re--calibration {\it did
not allow any longer} to forget the problem: the same theoretical RGB, for a
smaller age, provides a larger $\delta(B-V)$ and the theoretical RGs were
{\it too red}. Thus the ``new" ages required a change either in the
convection modelling, or a different tuning of the correlations colors-\Teff,
or both.
The situation is schematically shown in figure \ref{figalfa}, in which
I use an extreme difference in the distance modulus (0.25mag) to clarify the
problem. Suppose that color magnitude diagram was well fit by an isochrone
(open squares in figure \ref{figalfa}. An update of the distance modulus to
0.25mag longer (and the adjustment of the color by 0.06mag, in the range of
TO color uncertainties) provides now an age $\sim 6$Gyr younger, but it does
not allow to fit of the RGB. A good fit requires a {\it bluer} RGB.

The modellers have tried to solve the problem of the discrepancy between the
$\delta (B-V)$\ and the new distance scale in the following ways:
\begin{enumerate}
  \item they have {\it increased $\alpha$ to obtain again the fit}. On
theoretical grounds, there is no scientific basis in the assumption that the
$\alpha$\ in different stars should be the same as in the solar model, so
why not? This solution is adopted e.g. by Brocato et al. (1998) who discuss
at length the effect described here for the case of the GC M68)
and by Cassisi et al. 1998;
  \item some researchers have considered again models with solar $\alpha$,
but have {\it chosen the color \Teff relation in an appropriate way to
reproduce the giant colors} (it is generally possible to find good
justifications for this choice also). This is the approach by Salaris
and Weiss 1997, 1998: they adopt Buser and Kurucz (1978) colors for the
giants, which are bluer by several hundreths of magnitude than the more
recent ATLAS9 updated colors (see e.g. the comparison in figure 1 of Cassisi
et al. 1999).
In this way, the $\delta (B-V)$ between the TO and the RGB results {\it
smaller} and can fit GCs shapes with the new distance scale.
  \item there are a few attempts to {\it try different convection models},
which will generally not allow a ``perfect fit".
\end{enumerate}
If one adopts the solutions 1) or 2), it is important to remember that the
shape of the HR diagram loses any predictive power, as it has been fit
already assuming a distance scale: just as the ``spectacular
fits" of a few years ago produced a 15Gyr answer, present day fits will
produce a 10-12Gyr answer, but the quality of the fit has nothing to do
with the truth of the answer. {\sl A better way of posing the problem, when
an observer adopts a given set of tracks to infer the age of a new stellar
system, would be to say that the system shows the same age of -or that its
age differs from- the GC on which the track set has been more or less
explicitly calibrated}. It is not clear to me that even relative ages of GCs
of different metallicities can be inferred from the $\delta(B-V)$\ or
$\delta(V-I)$\ method, when we use a given MLT prescription which already is
tested on the HR diagrams to fit a given distance scale.

The 3rd solution is less misleading, and it could in the end allow progress
in the field, but it requires lots of work and maybe frustrating, as it
produces results not always in ``perfect agreement" with
observations\footnote{One of the reasons why we could get the hint of a
decrease in the GC ages two years before Hipparcos (Mazzitelli et al. 1995)
was that we used a convection models by which it was not possible to get a
fine tuning of the RG location.}. A few such attempts to overcome the MLT
are today available:
\begin{enumerate}
\item the FST (or Full Spectrum Turbulence) models, based on the Canuto and
Mazzitelli (1991) formulation, whose fluxes are in good agreement with
experimental data, and are computed using modern closures of the Navier
Stokes equations, and in which the scale length is assumed to be the distance
from the convective boundary. Models have been computed by Mazzitelli et al.
1995, D'Antona and Mazzitelli 1997, Silvestri et al. 1998. This formulation
of convection gives a different flavour to the HR diagram shape and it is
less tunable than the MLT, an advantage in terms of predictive power, but a
real failure if we want to obtain perfect fits. However, the
Silvestri et al. (1998) models, which differ from the previous of our group
mainly for the updated choice of color-\Teff\ transformations (Castelli
1998\footnote{see Kurucz website http://cfaku5.harvard.edu} versus Kurucz 1993),also provide a reasonable fit of the RGB as
shown in figures 2 and 6 --but notice also the discrepancy in the case of M30
in figure 5.
\item
Freytag and Salaris (1999) have calibrated the MLT $\alpha$\ by RHD models
based on grids of 2D hydrodinamic simulations by Ludwig et al. 1999.
Although numerical simulations are able to take into account only a
relatively small number of eddies for a realistic description of
turbulence, the Freytag and Salaris approach is an interesting novelty for
this field.
\item an incongruence of FST models and of models not adopting a plain MLT
description is that in any case they use till now grey boundary conditions,
and the colors are obtained through transformations based on MLT model
atmospheres. Models including, e.g., FST model atmospheres should be built
up to get selfconsistent colors (Kupka, Schmidt and D'Antona 1999);
\end{enumerate}

\section{The TO and upper MS location}
\label{toums}

The location of the TO is affected by many uncertainties in the input
physics, although not at the level of the RGB. If we wish to use the
theoretical MS colors to determine an age, we must shift vertically the
cluster HR diagram until it is superimposed to the MS of the observed
metallicity, and then we determine the age from the TO luminosity. The MS is
very steep in the TO region: a simple shift in color of the
theoretical MS by +0.02mag implies a determination of age smaller by 2Gyr,
not to talk about the possible uncertainty in the reddening. The main inputs
affecting the MS and TO location are the following:
\begin{enumerate}
 \item the convection description affects both the TO color and its shape
(see the comparison between the MLT based description and the FST models in
Mazzitelli et al. 1995 and D'Antona et al. 1997);
 \item The helium gravitational and thermal settlings (diffusion) affect
both the TO color and the age. A number of
models are available, starting from
Proffitt and Michaud 1991, up to D'Antona et al. 1997, Straniero et al. 1997
and Cassisi et al. 1998);
  \item the color - \Teff\ relation is affected by the convection treatment
in the atmosphere (cf. Kurucz 1993 versus Castelli 1998 models).
\end{enumerate}
Everything included, the absolute determination of the TO and upper MS
colors is uncertain by $\sim 0.05$mag, so that it is better not to rely on
colors for age determination.

I add a few words about the possible effect of helium diffusion. It is today
well settled that it is necessary to include the treatment of microscopic
helium diffusion to account for some details of the seismic Sun (Bahcall and
Pinsonneault 1995, Basu et al. 1996), but the evaluation of the diffusion
coefficients is difficult, and its application to models not always well
clear in the researchers description. The diffusion affects both the TO color
and the age: age reductions from 5-10\% to 20\% are found, and the TO color
may be affected up to 0.1mag in some models.
An important warning first issued by Deliyannis and Demarque (1991)
must be kept in mind: diffusion affects lithium nearly in the same way as
helium, thus: ``the properties of the Spite plateau in population II
severely restrict the amount of diffusion induced curvature that can be
tolerated in a lithium isochrone". In other words, the effect of ``too much"
diffusion would appear in a smaller lithium abundance for the hotter
population II stars, a fact which is not verified in the halo
subdwarfs, which show a remarkably flat lithium abundance
versus \Teff\ (the Spite and Spite 1982 plateau)\footnote{Here again we
attribute to GCs stars the same properties of the nearby subdwarfs. Actually,
the lithium behaviour at the TO of GCs might be a bit different than in the
field stars. Boesgaard et al. 1998 show that the M92 TO stars have a larger
scatter in lithium than field stars. The Spite's plateau must still be
confirmed by extensive GC stars observations, which are becoming possible
with the new generation telescopes.}. Chaboyer et al. 1992 show that an age
reduction up to 3Gyr (15\%) is in principle possible for GCs when diffusion
is included in the computation, but they find that the lithium isochrones
imply a maximum age reduction by 1Gyr ($\sim 7$\%).

In conclusion, also the theoretical TO -- upper MS color location is affected
by uncertainties by which the absolute age determination, again, can not be
known to better than $\pm$ several billion years.

\section{Location of the lower MS}

The ``double kink" shape of the low MS of GCs is due to the influence of the
interior physics on the structure of low mass stars. The appearence of the FK
is attributed mainly to the to the lowering the adiabatic gradient when the
$H_2$\ dissociation begins to be present in envelope (below $\sim 5000$K).
The SK is associated with the reaching of degeneracy in the interior
(D'Antona and Mazzitelli 1996). The shape of the low MS can be a powerful
tool, first to constrain the models, and then to constrain the GC parameters.

As first shown by Baraffe et al. 1995, when the formation of molecules begin
to be important in the stellar atmosphere (at $T_{\rm eff} \simlt 5000$K)
the grey atmospheric integration fails to give a good description for the
boundary conditions: in summary, it underestimates the
opacities and does not account for the opacity distribution with wavelength.
The net effect is that the grey integration provides much larger pressure
and density at the bottom of the atmosphere. In the interior, the temperature
gradient is the adiabatic gradient, so that finally the same central
conditions give a larger \Teff\ . Figure \ref{hrteo} shows in fact that, for
the same chemistry, non--grey models are redder by $\sim 0.06$mag with
respect to grey models. This also implies that the {\it metallicity} and
probably also the element to element ratios, are important to determine the
location of the FK. This is certainly a powerful tool, but also makes the
region below the FK very dependent on the model inputs.

Te EoS adopted for these low mass models is also an important ingredient.
It determines the {\it slope} of the region between the two kinks and,
together with the atmospheric integration, it influences the mass luminosity
relation, which is the most important input for the interpretation of
the luminosity functions of the MS in terms of mass function. There are
still substantial uncertainties close to the bottom of the main sequence
(see Montalban et al. 1999).

Figure \ref{hrteo} summarizes the uncertainties in the low MS location and
part of the uncertainties in the upper MS and TO locations. We see
that there is a region between $M_v = 6$\ and 7 at which the uncertainties
in color transformations, convection, diffusion, boundary conditions
seem to play almost no role: This region, then, is {\it the best to be used
as distance indicator for the MS}.

\section{Consistency of HB and low MS distance indicators}
\label{lowms}

The new HST data which have so much extended our knowledge on the low
luminosity part of the HR diagram  put an interesting problem: is there
consistency between the ``optical" traditional distance indicators for GCs
and the location and shape of the low MS?

We can check this idea by following this procedure: first we can fit the
optical data to the RR Lyrae (or HB) to derive a distance, and check the
reddening by controlling the MS location at $M_v \sim 6 -7$; then we adopt
the same reddening and distance modulus for the MS HST data. This allows us
to see if the MS and first kink location are consistently reproduced.

We show in the figures \ref{m92} and \ref{m30} the check on the low
metallicity ([M/H]=-2) clusters M92 and M30, finding excellent consistency.
The comparisons are equally good in the HST color bands (F555 and F814) and
in the transformed Johnson--Cousins bands $V$\ and $I$\ (in fact these HST
bands and the standard magnitudes are only marginally different). Thus, on
the one side the location of the FK provides a check of the distance, on the
other side we gain confidence in the good quality of the low mass models
including non grey boundary conditions.

Here again we must admit that this result is not {\it unique}: if we assume
for M30 a {\it short} distance modulus, say $(m-M)_0=14.5$, the general
agreement of the sequences would still be reasonable (and the age would
increase to $\sim 16$Gyr). Of course our HB models would not fit the HB
luminosities of the cluster, but we have agreed that even a small change in
the input physics may lead to less luminous HBs\footnote{Some further
indications on the choice of the ``best" distance can come from the
comparison of the observed and theoretical luminosity functions: D'Antona
(1998) and Silvestri et al. (1998) show that the non monotonic mass functions
derived, e.g. by Piotto et al. 1997, and by others, were mostly due to the
use of too short a distance modulus for the examined clusters.}.

\section{The distance through the fit to the lowest MS}
\label{secondk}

This approach has been applied by Reid and Gizis (1998) to the only cluster
for which the lowest main sequence, the post-SK region, is known,
namely NGC 6397 (figure 1). This part of the HR diagram is populated
by stars which are close to degeneracy in the interior, so they are close to
the minimum mass which can ignite hydrogen. The mass--luminosity relation
here is very steep, that is, the stars have practically all the same mass,
and the HR diagram location follows a constant radius line.
In addition, the degeneracy radius is not so much dependent on the details of
the structure, and this locus is very similar for all GC metallicities.
Thus, although the models are not well understood in detail here (Montalban
et al. 1999), if we have a reasonable sample of M subdwarfs with known
distances and accurate colors, we can fit the GC sequence to the nearby
sequence and get the distance modulus. The example of the fit by Reid and
Gizis (1998) is very interesting, and provides for NGC 6397 a not
unreasonable modulus $(m-M)_0=12.13 \pm 0.15$mag, again consistent with the
long distance moduli and short ages. We need a better definition of the
lowest MS through a larger sample of M subdwarfs, and more GC explored down
to the post-SK region, to make this distance indicator more useful.

\section{White dwarfs}
\label{wd}

There are by now three well defined WD sequences for GCs: NGC6752 (Renzini
et al. 1996), NGC6397 (King et al. 1998, see figure 1) and M4 (Richer et al.
1997). For the two latter clusters, the location of the WD sequence is
consistent with the cooling track for $M \simeq 0.5$\Msun\ by Wood (1995)
models transformed into the observed magnitudes by means of Bergeron et al.
(1995) model atmospheres (see Richer et al. 1997). The errors on the
reddening of the clusters and on the model observational colors are yet such
that we can not quantify better this general statement. The data of NGC6752
have been compared by Bragaglia et al. with a ``standard" sequence of field
WDs spectroscopically determined to be of mass $\simeq 0.5$\Msun (from the
list of Bragaglia et al 1995). They obtain a ``short" distance modulus, and
an age of $\sim 15$Gyr (even assuming a large $\alpha$--enhancement). This
result is then marginally discrepant from the others we have quoted so far.

On the other hand, from Wood (1995) models we see that $\Delta M_v/ \Delta
M_{wd} \simeq 2.5$. In other words, a small uncertainty (by 0.1\Msun) in the
mass determination fo the field WD sample produces a noticeable difference
(by 0.25mag) in the WD sequence location, which results in an age difference
close to 4Gyr. Do we know the spectroscopic masses within 0.1\Msun?
We know that the spectroscopic masses of {\it helium atmosphere} WDs are
highly uncertain, some uncertainty surely weighs also on the DA type WDs. It
is also possible that the field WDs differ from the GC WDs in other ways: the
environment in which they are born is very different and may imply, e.g.,
different accretion rates, much larger in the disk than in the GC, which is
devoid of gas and dust. Although the sedimentation of metals is very
efficient in these high gravity stars, it is well possible that some residual
effect from accretion affects the stellar radius.

\section{Summary}
I conclude with the following short summary:

-- {\it Morphological} fits including the RGB are meaningless in terms
of age determination;

-- The traditional best theoretical {\it distance scale} still mostly
relies on HB models, and on the MS colors at $M_v \sim 6 - 7$.

-- the {\it small} versus {\it large} ages only require a difference
of $\simeq 0.25$mag of distance modulus;

-- The uncertainties in HB, TO, MS, WDs sequences location are all of
the same order of magnitude: namely: $\sim 0.25$mag;

-- then we can make no firm choice between 10-12 and 14-16Gyr age;

-- however {\it most} recent theoretical and observational results,
including the new constrains on distance by the very low luminosity stars,
all point towards the smaller range of ages.

%
\section*{Acknowledgements}
I warmly thank the organizers of the Colloquium, and especially Arlette
Noels, first for inviting me, and second for holding a perfect scientific
and logistic meeting.

I also thank my coworkers and friends Vittoria Caloi and Italo Mazzitelli,
together with Josefina Montalban, Paolo Ventura, and Fabio Silvestri for
always useful exchange of ideas and for the work which has made this review
possible. I acknowledge all the useful data and lively information from
F. Allard, A. Cool, I. King, G. Piotto, H. Richer and M. Salaris.

%
 
\beginrefer
\refer Andreuzzi et al. 1999, submitted to A\&A

\refer Bahcall, J.N., \& Pinsonneault, M. 1995, Rev. Mod. Phys. 67, 781

\refer Baraffe I., Chabrier G., Allard F., Hauschildt P., 1995 ApJ 446, L35

\refer Baraffe I., Chabrier G., Allard F., Hauschildt P., 1997 A\&A 327,
1054 (BCAH97)

\refer Bergeron, P. Wesemael, F. \& Beauchamp, A. 1995, PASP 107, 1047

\refer Basu, S., Christensen Dalsgaard, J., Schon, J., Thompson, M.J. 1997,
ApJ 460, 164

\refer Bergbusch, P.A., \& VandenBerg, D.A. 1992, ApJS 81, 163

\refer Boesgaard, A.M. et al. 1998, ApJ 493, 206

\refer Bragaglia, A., Renzini, A., Bergeron, P. 1995, ApJ 443, 735

\refer Buonanno, R., Corsi, C. \& Fusi Pecci, F. 1985 A\&A 145, 97

\refer Buser, R., Kurucz, R. 1978, A\&A 70, 555

\refer Caloi, V., D'Antona, F., \& Mazzitelli, I. 1997, A\&A 320, 823

\refer Canuto, V.M., \& Mazzitelli, I. 1991, ApJ 370, 295

%

\refer Cassisi, S., Castellani, V., Degl'Innocenti, S., Weiss, A. 1998,
A\&AS 129, 267

\refer Cassisi, S., Castellani, V., Degl'Innocenti, S., Salaris, M., Weiss,
A. 1999, A\&AS 134, 103

\refer Castelli, F. 1998, in ``Views on distance indicators", Mem.S.A.It.
69, 165

\refer Chaboyer, B. 1995, ApJ 444, L9

\refer Chaboyer, B., Deliyannis, C.P., Demarque, P., Pinsonneault, M.H.,
Sarajedini, A. 1992, ApJ 388, 372

\refer Chaboyer, B. \& Kim, Y.--C. 1995, ApJ 454, 767

\refer Chaboyer, B., Demarque, P., Kernan, P.J., Krauss, L.M. 1996, Science
271, 957

\refer Chaboyer, B., Demarque,P., Kernan, P.J., Krauss, L.M. 1998, ApJ 494,
96

\refer Chieffi, A., Straniero, O., Salaris, M. 1995, ApJL 445, L39


\refer Cool, A.~M. 1997, ``Binary Stars Below the Turnoff in Globular Cluster
Color--Magnitude Diagrams'' in {\it Advances in Stellar Evolution,} eds.
R.~T.~Rood and A. Renzini (Cambridge: Cambridge U. Press), p.~191

\refer D'Antona F., 1995,in The Bottom of the Main sequence -- and beyond,
WSO Astrophysics Symposia (Springer) Ed. C. Tinney p.13

\refer D'Antona F. 1998, in ``The Stellar Initial Mass Function", ed.
G.Gilmore \& D. Howell ASP Conference Series, (San Francisco: ASP), vol. 142,
p.~157

\refer D'Antona F., Caloi V., Mazzitelli I., 1997, ApJ 477, 519

\refer D'Antona F., Mazzitelli I., 1996, ApJ 456, 329

\refer Deliyannis, C. \& Demarque, P. 1991, ApJ 379, 216

\refer Feast, M.W. \& Catchpole, R.M. 1987, MNRAS 286, L1

\refer Fernley, J. Barner, T.G., Skillen, I., Hawley, S.L., HAnley, C.J.,
Evans, D.W., Solano, E., Garrido, R. 1998, A\&A 330, 515

\refer Freytag, B. \& Salaris, M. 1999, ApJL 513, L49

\refer Gratton R.G., Fusi Pecci F., Carretta E., Clementini G., Corsi C.E.,
Lattanzi M., 1997, ApJ 491, 749

\refer Groenewegen, M.A.T. \& Salaris, M. 1999, A\&A 348, L33

\refer Kaluzny, J. 1997, A\&AS 122,1

\refer King I.R., Anderson J., Cool A.M., Piotto G., 1998, ApJ 492, L37

\refer Kupka, F., Schmidt, M. \& D'Antona, F. 1999, in preparation


\refer Kurucz, R.L. 1991, in: Stellar Atmospheres: Beyond the
Classical Models, L. Crivellari, I. Hubeny, D. G. Hummer eds., NATO ASI
Series (Dordrecht: Kluwer), p. 441

\refer Kurucz, R.L. 1993, ATLAS9 Stellar Atmosphere Programs and 2 km/s grid
   (Kurucz CD-ROM No 13)

\refer Layden, A.C., Hanson, R.B., Hawley, S.L., Klemola, A.R., Hanley, C.J.
1996, AJ 112, 2110

\refer Ludwig, H.-G., Freytag, B. \& Steffen, M. 1999, A\&A 346, 111

\refer Mazzitelli, I., D'Antona, F. \& Caloi, V. 1995, A\&A
302, 382 (MDC)

\refer Montalban, J., D'Antona, F., Mazzitelli, I. 1999, submitted to A\&A

\refer Piotto et al. 1990 ApJ 350,662

\refer Piotto G., Cool A. \& King I.R., 1997, AJ, 113 1345

\refer Pont, F., Mayor, M., Turon, C. \& Vandenberg, D. A. 1998, A\&A 329,
87

\refer Proffitt, C.R. \& Michaud, G. 1991, ApJ 371, 584

\refer Reid I.N., 1997, AJ 114, 161

\refer Reid I.N., \& Gizis, J.E. 1998, AJ 116, 2929

\refer Richer, H. et al. 1997, ApJ 484, 741

\refer Rogers F.J., Swenson F.J., Iglesias C.A., 1996, ApJ 456, 902

\refer Salaris, M., Chieffi, A., and Straniero, O. 1993, ApJ 414, 580

\refer Salaris, M., Degl'Innocenti, S., and Weiss, A. 1997, ApJ 479, 665

\refer Salaris, A., Weiss, A. 1997, A\&A 327, 107

\refer Salaris, A., Weiss, A. 1998, A\&A 335, 943

\refer Sandage, A., 1993 AJ 106, 703

\refer Silvestri F., Ventura, P., D'Antona F., Mazzitelli I., 1998, ApJ 509,
192

\refer Spite, F. \& Spite, M. 1982, A\&A 115, 357

\refer Stetson, P.B. \& Harris, W.E. 1988, AJ 96, 909

\refer Straniero, O., Chieffi, A., Limongi, M. 1997, ApJ 490, 425

\refer VandenBerg, D.A. 1983, ApJS 51, 29

\refer VandenBerg, D.A. 1992, ApJ 391, 685.

\refer VandenBerg, D.A., \& Bell, R.A. 1985, ApJS 58, 561

\refer VandenBerg, D.A., Bolte M., \& Stetson, P.B. 1990, AJ 100, 445

\refer VandenBerg, D.A., Stetson, P.B. \& Bolte M. 1996, ARAA 34, 461

\refer Walker, A.R. 1992, ApJ 300, L81


\refer Wood, M.A. 1995, in ``White Dwarfs", ed. D. Koester \& K.Werner
(Berlin:Springer), p.41

\endrefer           
\newpage

\begin{figure}
  \caption{Composite HR diagram for the stars in NGC6397. The upper part is
adapted from Kaluzny 1997, the giants and turnoff data are from Cool 1997,
and the main sequence down to its low end, plus the white dwarfs are from
King et al. 1998.} \label{f1}
\end{figure}

\begin{figure}
  \caption{Main evolutionary and structural phases in the HR diagram:
WD: 0.5\Msun CO cooling track from Wood
1995; MS: from Baraffe et al. 1997 (BCAH97) for [M/H]=-1.5; TO+RGB: 12Gyr
isochrone from Silvestri et al. 1998; HB from Caloi et al. 1997; AGB is
schematized.} \label{f2}
\end{figure}

\begin{figure}
\caption{In recent years, very small variations in the HB model luminosity
and in the TO luminosity for fixed age have produced a relatively huge
effect on the age determination. The ``OLD" lines represent a typical
location in absolute visual magnitude of HB (top) and TO (bottom) models in
old stellar models. A small increase in the HB luminosity (by $\sim 0.07$mag)
has been due to the update in the interior EoS of HB models (Rogers et al.
1996), another increase by $\sim 0.07$mag has been due to a small increase in
the core mass at the Helium flash (see Caloi et al. 1997). At the same
time, the TO luminosity at a given age has been slightly decreased by updates
in the EoS (again it is an effect of the Rogers et al. 1996 EoS, found by
Chaboyer and Kim 1995 and D'Antona et al. 1997) and another small decrease
(again by $\sim 0.07$mag can be attributed either to the convection model (a
{\it shape} effect due to the Canuto and Mazzitelli 1991 treatment of
convection, in Mazzitelli et al. 1995 and D'Antona et al. 1997), and/or to
the influence of helium diffusion (gravitational settling). In total, the
new models give a $\Delta V$ between TO and HB which is $\sim$.27mag larger,
and therefore a given $\Delta V$ corresponds to an age $\sim 4$Gyr lower. }
\label{deltav} \end{figure}

\begin{figure}
\caption{A schematic view of the problem with the RGB location is shown:
a good fit with a distance scale which would provide 16Gyr age for a
cluster, is no longer OK if the distance modulus is increased. Here we show
an increase by 0.25mag to enphasize the problem, but the situation is
similar if the shift is smaller. In practice, a more efficient convection,
or a different color -- \Teff\ relation for the RGB, is necessary to fit
again the RGB, as shown by the dotted line. In this sense, all set of tracks
have their convection treatment plus color-\Teff\ relations ``calibrated" on
a distance scale. }\label{figalfa} \end{figure}

\begin{figure}
\caption{In the plane $M_v$\ versus $V-I$\ the effect of some important
physical inputs on the HR diagram location is shown. The low main sequence
location is about 0.06mag redder when models are computed with non grey
boundary conditions (Montalban et al. 1999), with respect to the grey
atmosphere models (Silvestri et al. 1998). Luckily, the color-\Teff relations
for the low main sequence are not so dependent on the employed
transformations, as shown by the location of the Silvestri et al. models
transformed via the Allard and Hauschildt 1997 NextGen models (AH) or with
Castelli 1998 models. On the contrary, the transformations produce a
small shift of the turnoff color, by $\sim 0.03$mag. At the TO, also the
treatment of convection affects the isochrone colors. The less critical
models are those of masses $\sim 0.6 - 0.65$\Msun\ and $M_v \sim 6-7.5$,
where the non-grey boundary conditions are not yet important, and where the
effect of superadiabaticity and also the effect of diffusion are minimum.
This is then the best region where to require a fit of the theoretical and
observational main sequence.}\label{hrteo} \end{figure}

\begin{figure}
  \caption{Composite HR diagram for M92. The left part shows the fit of the
optical ground based data in $V$\ versus $B-V$, including the HB (from
Buonanno et al. 1985) and the RGB+TO+MS data (from Stetson and Harris 1988).
The right part are the Andreuzzi et al. HST data shifted at the
distance of and dereddened as the optical data. We adopt $E(V-I)=1.3E(B-V)$\
based on Allard and Hauschildt model atmospheres. The isochrones are from
Montalban et al. (1999) for the indicated chemistry.
} \label{m92}
\end{figure}

\begin{figure}
  \caption{The same comparison is shown for M30. The optical data (from
Piotto et al. 1990 for the MS and TP, and from Buonanno et al. 1985 for the
RGB and HB) are fitted by assuming a distance modulus of 14.7 and
E(B-V)=0.04. The same choice provides a good reproduction of the HST data for
the MS including the first kink. The dashed line here and in figure
\ref{m92} is the corresponding
12Gyr isochrone from D'Antona et al. 1997 (DCM), which employed Kurucz
(1993) color transformations and grey atmosphere models.
} \label{m30}
\end{figure}
\vfill

\end{document}